\documentclass[bibnotes,prl,twocolumn,superscriptaddress]{revtex4}%
\usepackage{epsfig,dsfont,amssymb,amsmath,amsthm,amsfonts,amsbsy,mathrsfs}
\usepackage{graphicx}
\usepackage{bm}
\usepackage{epstopdf}

\begin{document}
\title{Transmission Nonreciprocity in a Mutually Coupled Circulating Structure}
\author{Bing He}
\email{binghe@uark.edu}
\affiliation{Department of Physics, University of Arkansas, Fayetteville, AR 72701, USA}
\author{Liu Yang}
\affiliation{College of Automation, Harbin Engineering University, Harbin 150001, China}
\author{Xiaoshun Jiang}
\affiliation{National Laboratory of Solid State Microstructures, College of Engineering and Applied Sciences, School of Physics,
Nanjing University, Nanjing 210093, China}
\author{Min Xiao}
\email{mxiao@uark.edu}
\affiliation{Department of Physics, University of Arkansas, Fayetteville, AR 72701, USA}
\affiliation{National Laboratory of Solid State Microstructures, College of Engineering and Applied Sciences, School of Physics,
Nanjing University, Nanjing 210093, China}

\begin{abstract}
Breaking Lorentz reciprocity was believed to be a prerequisite for nonreciprocal transmissions of light fields, so the possibility 
of nonreciprocity by linear optical systems was mostly ignored. We put forward a structure of three mutually coupled microcavities 
or optical fiber rings to realize optical nonreciprocity. Although its couplings with the fields from two different input ports are constantly equal, 
such system transmits them nonreciprocally either under the saturation of an optical gain in one of the cavities or with the asymmetric couplings of the circulating fields in different cavities. The structure made up of optical fiber rings can perform nonreciprocal transmissions as a time-independent linear system without breaking Lorentz reciprocity. Optical isolation for inputs simultaneously from two different ports and even approximate optical isolator operations are implementable with the structure.
\end{abstract}

\maketitle


The availability of the approaches other than those based on Faraday effect \cite{mag,mag1,mag3} is essential to realizing integrated circuits 
of optical nonreciprocity. ``Optical nonreciprocity" here refers to the phenomena that the transmission of a field with certain frequency or bandwidth
from port $P_a$ to port $P_b$ in a circuit is asymmetric with its reversed transmission from $P_b$ to $P_a$, being different from ideal ``optical isolator" \cite{criteria} that completely blocks the field from one of the ports, irrespective of its modal contents and polarizations. 
Synthetic magneto-optical effects, such as those of spatio-temporal modulations \cite{match1, al1, al2, al3, aa0, al4, aa, lattice} and optomechanics \cite{oms1,oms2,oms3}, were primarily considered as the replacement for Faraday effect. Numerous other methods \cite{b2, b3, b4, b5, b6, b7, b8, bragg, wen, jiang, match, atom1, atom2,o1, o2, band, gain} were also found for the purpose.

To propagate asymmetrically, light fields should be under an effect depending on their propagation directions.
For example, the momenta of photons should satisfy a phase-matching condition for interband transitions \cite{match1} or in parametric processes \cite{match}. Or else, nonreciprocal transmissions can occur by unequal couplings of a system to the inputs from two directions. Such examples include the ones making use of the optical gain saturated disparately under the fields unequally coupled from two waveguides \cite{gain, wen, jiang}, as well as certain atomic systems non-identically coupled to photons from different directions \cite{atom1, atom2}. One question is whether there exists a nonreciprocal structure that is only made of isotropic media and can couple to the inputs from two different sides identically. It is also a widely held notion that the relation derived from Lorentz reciprocity theorem \cite{criteria, book} must be broken for achieving optical nonreciprocity. Whether this assumed restriction in designing nonreciprocal devices can be removed for certain systems, which well perform nonreciprocal transmissions, is fundamentally meaningful to understanding optical nonreciprocity. 

In this Letter we provide the definite answers to these questions, using the mutually coupled circulating structure (MCCS) in Fig. 1(a). 
Developed from the widely studied {\cal PT}-symmetric optical systems \cite{gain, pt1, pt2, pt4, pt3, pt5, pt7, pt8} by adding one more passive component (cavity 2), such system can be constructed with three mutually coupled microresonators \cite{3unit1, 3unit2} or optical fiber rings. 
Its relevant features are illustrated in what follows.

Inside each cavity of the MCCS, a circulating field mode couples to the reversely circulating ones in the neighboring cavities. Any external drive thus creates three pairs of circulating modes (one clockwise (CL) and one counter-clockwise (CCL) in each cavity) through the mutual couplings of the topological structure. In the frame rotating with the cavities' resonance frequency, the dynamical equations for the coupled cavity modes read
\begin{subequations}
\begin{align}
\dot{a}_1^+&=-\gamma_1 a_1^+-iJ_{12}a_2^--iJ_{13}a_3^-+\sqrt{2\kappa_e}S_{i,f},\\
\dot{a}_1^-&=-\gamma_1 a_1^--iJ_{12}a_2^+-iJ_{13}a_3^+,\\
\dot{a}_2^+&=-\gamma_2 a_1^+-iJ_{12}a_1^--iJ_{23}a_3^-,\\
\dot{a}_2^-&=-\gamma_2 a_2^--iJ_{12}a_1^+-iJ_{23}a_3^+,\\
\dot{a}_3^+&=(g(t)-\gamma_3) a_3^+-iJ_{13}a_1^--iJ_{23}a_2^-+\sqrt{2\kappa_e}S_{i,b},\\
\dot{a}_3^-&=(g(t)-\gamma_3) a_3^--iJ_{13}a_1^+-iJ_{23}a_2^+,
\end{align}
\end{subequations}
where ``$+(-)$" represents a CL (CCL) mode. Besides the dissipation rate $\gamma_k$ ($k=1,2$ and $3$) and amplification rate $g(t)$, 
the main parameters are the coupling coefficients $J_{ij}$ ($i,j=1,2$ and $3$ for $i\neq j$) determined by the distances between two microcavities or by the designs of fiber couplers. The MCCS enjoys richer dynamical properties than the typical PT-symmetric systems---as in Fig. 1(b1) the system will undergo the transition from instability to stability and vice versa, as the couplings of cavity 2 to the two other cavities change. This distinguishes the structure from another topology of linking the components into a string \cite{string1, string2}.

\begin{figure}[t!]
\vspace{0cm}
\centering
\epsfig{file=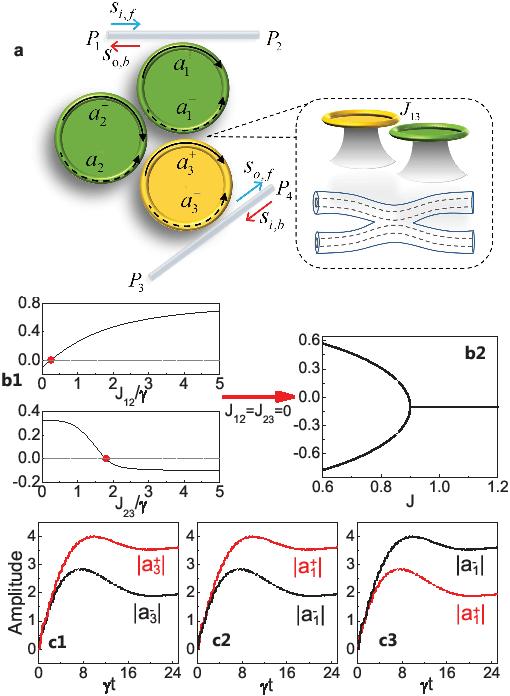,width=1.0\linewidth,clip=}
{\vspace{-0.4cm}\caption{Mutually coupled circulating structure and its dynamical properties. (a): The schematic of the structure, which can be constructed either with coupled microcavities or with coupled optical fiber rings. Cavity 3 (the yellow one) carries gain medium. The concerned nonreciprocity refers 
to the realized phenomena that the identical $S_{i,f}$ and $S_{i,b}$ induce the unequal $S_{o,f}$ and $S_{o,b}$. (b1): An example of the maximum real part of the eigenvalues of such a coupled system neglecting gain saturation, which varies with the couplings $J_{12}$ and $J_{23}$, respectively. The transitions between instability (vertical axis value $>0$) and stability (vertical axis value $<0$) occur at the marked points. (b2): The reduction to the dynamics of a PT-symmetric system when cavity 2 is detached, having the real parts of the eigenmodes to merge at an exceptional point. 
(c1): An example of the evolved field modes contributing to the outputs, due to a drive at $P_1$. (c2): The corresponding field modes
due to a same drive at $P_4$. (c3): The corresponding field modes by placing the drive at $P_3$. The plot for ``$+(-)$" mode in (c1) and (c2) 
swaps the position with that for ``$-(+)$" mode in (c3).}}
\vspace{-0.7cm}
\end{figure}

\begin{figure}[b!]
\vspace{-0cm}
\centering
\epsfig{file=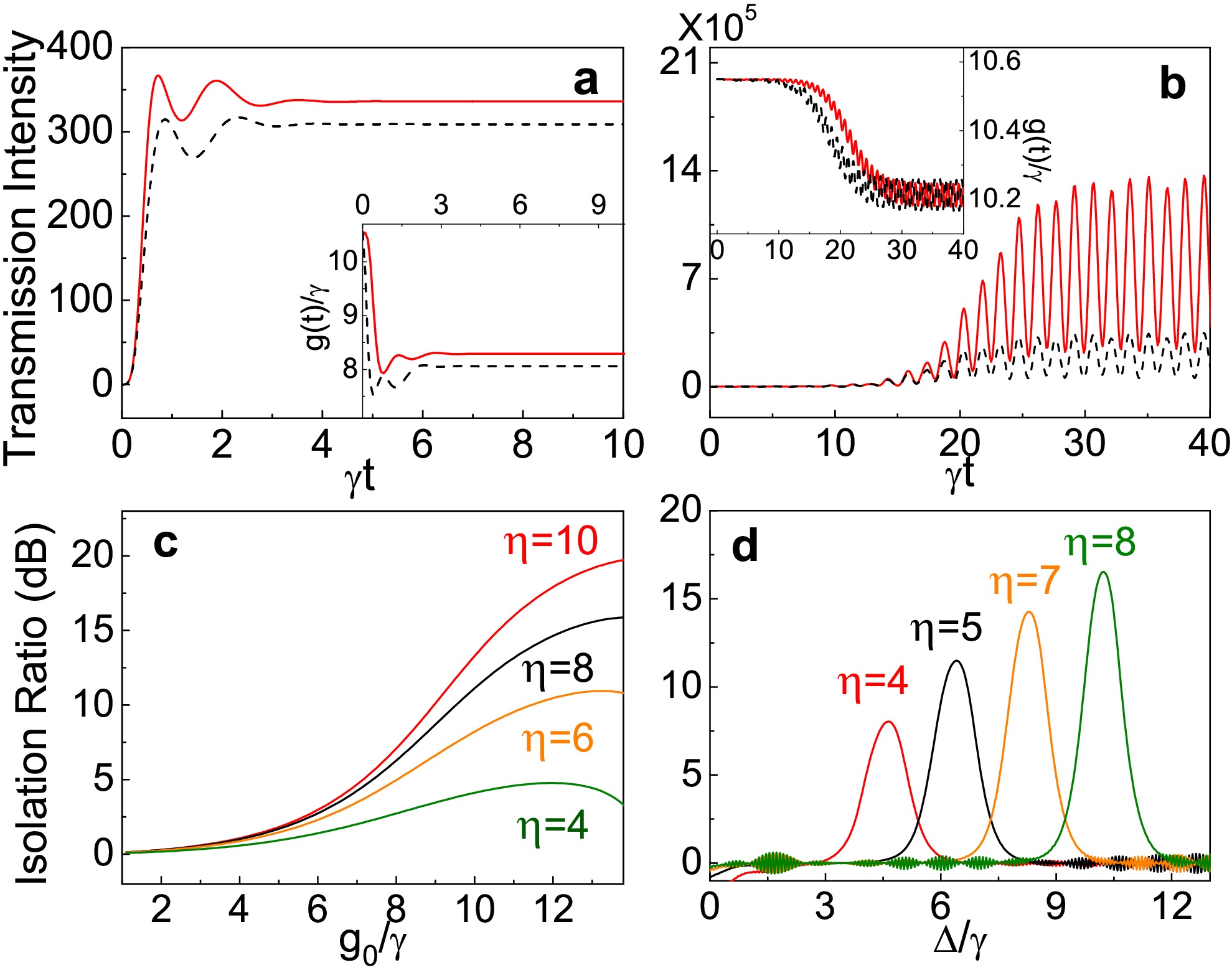,width=1.0\linewidth,clip=}
{\vspace{-0.4cm}\caption{Transmission nonreciprocity due to gain saturation. Here the relative parameters are chosen as $\gamma_1=\gamma_2=\gamma$, $\gamma_3=9.72\gamma$, $J_{ij}=2.2\gamma$, which are converted from a set of experimental parameters in \cite{gain}. (a) and (b): The evolutions of the forward intensity $|S_{o,f}|^2/(2\kappa_e)=|a_{3,f}^-|^2$ (solid curve) and backward intensity $|S_{o,b}|^2/(2\kappa_e)=|a_{1,b}^-|^2$ (dashed curve), with $g_0=10.55\gamma$, $\Delta=0$ in (a) and $\Delta=5\gamma$ in (b). These dimensionless intensities scale with the transmitted field powers. 
The dimensionless saturation intensity, which is determined by the scattering cross-section and lifetime of the dopant, and the effective volume of the field inside cavity, is given as $I_0=1.33\times 10^3$ in (a) and $1.33\times 10^7$ in (b). 
The inserted figures are about the corresponding gain rates. (c) and (d): The relations between isolation ratio and system parameters for the evolved time-independent steady modes by choosing $g_0=\gamma_3$ and the $I_0$ in (a), with different ratios $\eta=J_{23}/J_{12}$. In (c) the detuning $\Delta/\gamma$ is set to be $\eta$. The drive amplitude in these examples is $E_f=E_b=100\gamma$.}}
\vspace{0cm}
\end{figure}

Unlike many other systems for implementing nonreciprocal transmissions, the couplings of the MCCS to the forward input $S_{i,f}$ 
and the backward input $S_{i,b}$ are always the same with a fixed coupling constant $\kappa_e$. Relative to a forward drive 
at port $P_1$, we mainly consider a corresponding backward drive with the identical intensity from $P_4$, while the backward one can be applied on $P_3$ too.
Under a constant optical gain, the backward input from $P_3$ also gives the same output amplitude $S_{o,b}=\sqrt{2\kappa_e}a_{1}^-$ at $P_1$ 
as the forward one $S_{o,f}=\sqrt{2\kappa_e}a_{3}^+$ at $P_3$. Such relations of the outputs, partially due to Lorentz reciprocity, 
are displayed in Figs. 1(c1)-1(c3).

\begin{figure*}[t!]
\vspace{-0cm}
\centering
\epsfig{file=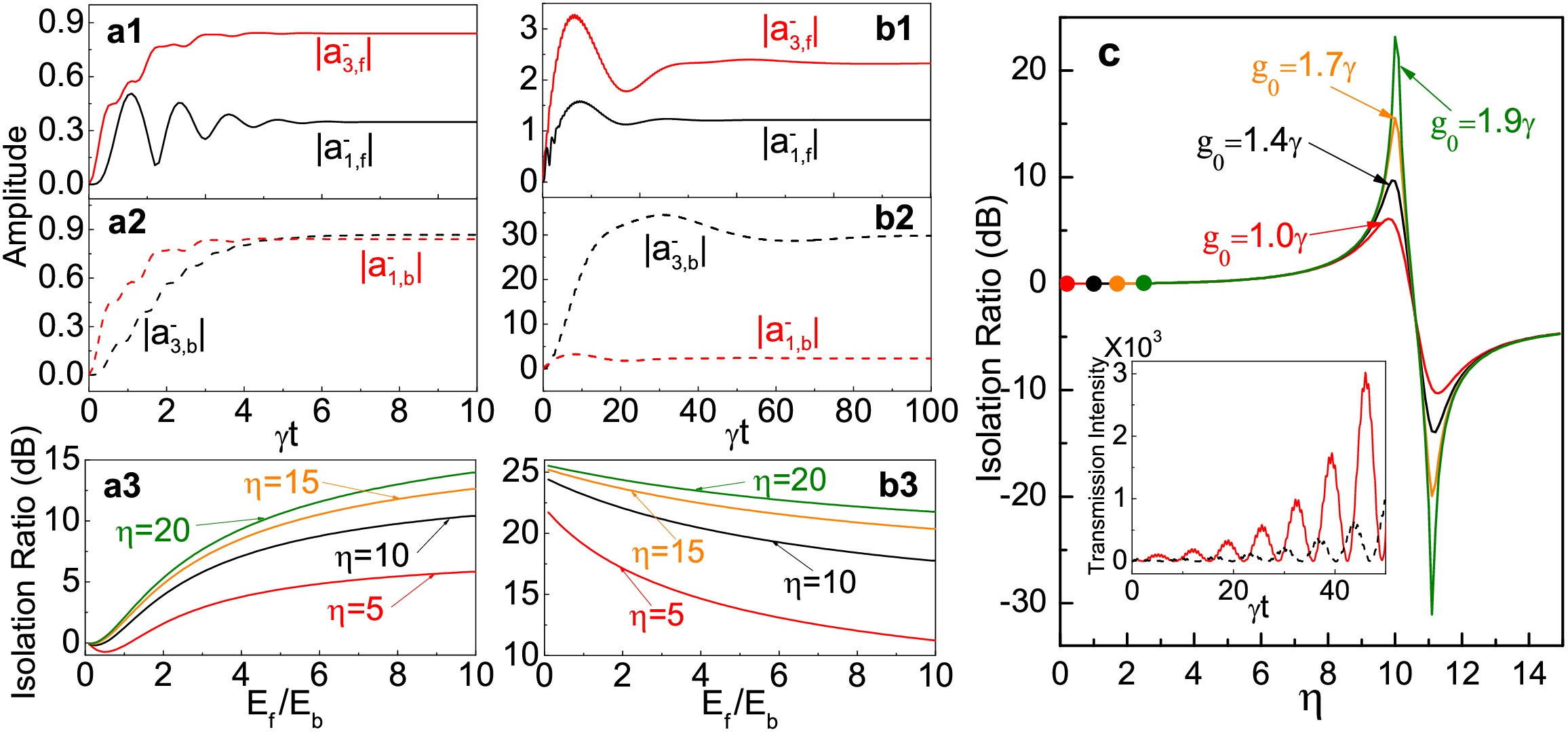,width=0.7\linewidth,clip=}
{\vspace{-0.4cm}
\caption{Optical isolation for two simultaneous inputs. (a1)-(a2): The induced field modes by the forward and backward drive, respectively. Here $g_0=0$, $\Delta=5\gamma$, $J_{12}=J_{13}=\gamma$, $E_f=E_b=10\gamma$, and $\eta=5$. $\gamma$ is the damping rate of the three cavities. (a3): The isolation ratios for two opposite drives acting together on the system, with $g_0=0$ and $\Delta=20\gamma$. (b1)-(b3): The corresponding results to those in (a1)-(a3), with a difference of $g_0=1.9\gamma$. (c): The relation between optical isolation ratio and $\eta$, 
where the finally time-independent field modes are obtained under any pair of identical and simultaneously acting inputs with
$\Delta=10\gamma$. The inserted figure illustrates an example of unstably evolving $|a_{3,f}^-(t)+a_{3,b}^-(t)|^2$ (solid) and 
$|a_{1,f}^-(t)+a_{1,b}^-(t)|^2$ (dashed), with $\eta=0.5$ on the left of the marked dot on the curve of $g_0=1.4\gamma$. Gain saturation is neglected here.}}
\vspace{-0.6cm}
\end{figure*}

A convenient tool to describe the relevant properties is scattering matrix, which takes the symmetric form,
\begin{eqnarray}
\hat{S} =\left(
\begin{array}{cccccc}
S_{1+}^{1+} & S_{1-}^{1+} & S_{2+}^{1+} & S_{2-}^{1+} & S_{3+}^{1+} & S_{3-}^{1+} \\
S_{1-}^{1+} & S_{1-}^{1-} & S_{2+}^{1-} & S_{2-}^{1-} & S_{3+}^{1-} & S_{3-}^{1-}\\
S_{2+}^{1+} &  S_{2+}^{1-} & S_{2+}^{2+} & S_{2-}^{2+} & S_{3+}^{2+}& S_{3-}^{2+}\\
S_{2-}^{1+} & S_{2-}^{1-} & S_{2-}^{2+} & S_{2-}^{2-} & S_{3+}^{2-} & S_{3-}^{2-}\\
S_{3+}^{1+} & S_{3+}^{1-} & S_{3+}^{2+} & S_{3+}^{2-} & S_{3+}^{3+} & S_{3-}^{3+}\\
S_{3-}^{1+} & S_{3-}^{1-} &S_{3-}^{2+} & S_{3-}^{2-} & S_{3-}^{3+} & S_{3-}^{3-}
\end{array}
\right),
\end{eqnarray}
from Eqs. (1a)-(1f). Rewriting a single-frequency input as $\sqrt{2\kappa_e}S_{i,f(b)}=E_{f(b)}e^{i\Delta t}$, where $\Delta$ is the detuning with respect 
to cavities' resonance frequency, one sees that the transmission reciprocity between $P_1$ and $P_4$ manifests as $S_{3-}^{1+}E=S_{3+}^{1-}E$, when two inputs with $E_f=E_b=E$ act separately on the system with a constant $g$.
It is the consequence of a dynamical symmetry of the structure, which exhibits as $S_{3-}^{1+}=S_{3+}^{1-}$,
the equality of the scattering matrix element $(16)$ with $(25)$ and $(61)$ with $(52)$ (the first and second number represent the matrix element's 
row and column, respectively), in addition to the overall symmetry $(ij)=(ji)$ of the scattering matrix.

The transmission reciprocity can be easily lost under gain saturation. The actual gain rate
$g(t)=g_0/\big(1+|a_3^+(t)+a_3^-(t)|^2/I_0\big)$ becomes stable with time according to the saturation intensity $I_0$ of the given gain medium, as in Figs. 2(a) and 2(b). With the different gain rates $g(t)$ in Eqs. (1a)-(1f) for the inputs $S_{i,f}$ and $S_{i,b}$ that are 
turned on separately, the nonreciprocal transmissions between ports $P_1$ and $P_4$ exist. The trick to implement such nonreciprocity 
is that, through the mutual couplings, the identical $S_{i,f}$ and $S_{i,b}$ induce unequal $|a_{3,f}^+(t)+a_{3,f}^-(t)|^2$ and $|a_{3,b}^+(t)+a_{3,b}^-(t)|^2$ in the active component. 
The transmission nonreciprocity is measured by the log-ratio $10\log_{10}(|a_{3,f}^-|^2/|a_{1,b}^-|^2)$, because the concerned forward and backward outputs are proportional to $a_3^-$ and $a_1^-$, respectively. Its relations with the system parameters are displayed in Figs. 2(c) and 2(d), where $\eta=J_{23}/J_{12}$ is an important factor adjusted by the couplings of cavity 2 with the two others. Rayleigh scattering \cite{rayleigh, thesis} 
that couples a CL mode with a CCL mode inside each cavity exists in realistic systems, and its effect is discussed in Sec. I of Supplementary Material \cite{supp}. 

To use the MCCS as an approximate optical isolator, a necessary condition is the highly asymmetric transmission contrast 
under the forward and backward drives together \cite{match, nonlinear}. The isolation of two simultaneous inputs is measured by the logarithmic ratio
\begin{eqnarray}
10\log_{10}\left |\frac{a_{3,f}^-+a_{3,b}^-}{a_{1,f}^-+a_{1,b}^-}\right|^2=10\log_{10}\left |\frac{S_{3-}^{1+}E_f+S_{3-}^{3+}E_b}{S_{1-}^{1+}E_f+S_{3+}^{1-}E_b}\right|^2,
\end{eqnarray}
which includes the contributions from the reflected fields of the simultaneous forward input $E_f$ and backward input $E_b$.
Optical isolation (a nonzero ratio of the above when $E_f=E_b$) immediately appears after breaking the symmetry of identical coupling between 
the cavities (having unequal distances between cavities so that $\eta\neq 1$), 
and it exists even without gain [Figs. 3(a1) and 3(a2)] and becomes more significant under gain [Figs. 3(b1) and 3(b2)]. 
Figs. 3(a3) and 3(b3) show that the associated isolation ratios defined in Eq. (3) stabilize with increased $E_f/E_b$. Moreover, they have two groups of peak values with the opposite signs, one of which is around $\eta=\Delta/\gamma$ due to a transmission resonance. In Fig. 3(c) showing this feature, the left of the marked dot on each curve is the unstable regime described in Fig. 1(b1). The outputs in the unstable regime are still asymmetric, as seen from an example of the time evolutions of their intensities.

Another symmetry breaking leads to a realization of nonreciprocity without gain saturation by the structure made up of optical fiber rings. 
The coupler illustrated in Fig. 4(a) capitalizes on directional macrobending loss \cite{macrobend}, so that more light from one direction couples into another fiber than from the reverse direction. One choice as in Fig. 4(b) is that, through the coupler, mode $a_1^+$ in cavity 1 and mode $a_3^-$ in 
cavity 3 couple more strongly than the other pair $a_1^-$ and $a_3^+$, to replace the coefficient $J_{13}$ in Eqs. (1a) and (1f) [Eqs. (1b) and (1e)] by $J_{13}+\epsilon$ ($J_{13}-\epsilon$). The couplings via such bending radiation are rather weak. 
To implement an optical isolator operation, which requires that both the reflection $a_{1,f}^-=S_{1-}^{1+}E_f$ of the forward input and the backward transmission $a_{1,b}^-=S_{3+}^{1-}E_b$ be highly suppressed while the forward transmission $a_{3,f}^-=S_{3-}^{1+}E_f$ keep to be significant, 
one can lower the coupling constants $J_{12}$ and $J_{23}$ for cavity 2 to the order of $J_{13}\pm\epsilon$, as illustrated by the example in Figs. 4(c1)-4(c2). More details of this mechanism are in Sec. II of Supplementary Material \cite{supp}.

\begin{figure}[t!]
\vspace{-0cm}
\centering
\epsfig{file=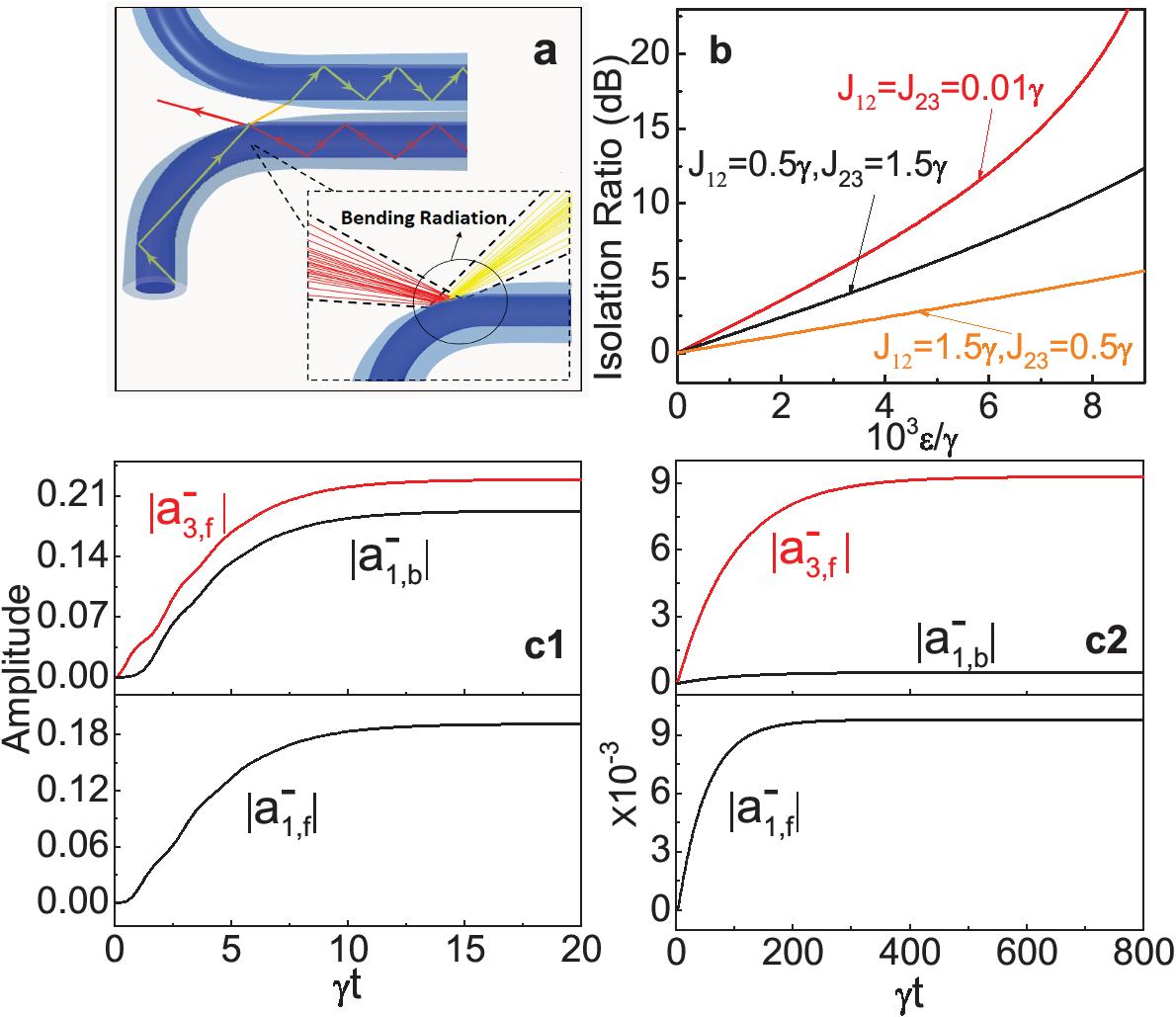,width=1\linewidth,clip=}
{\vspace{-0.5cm}\caption{Optical nonreciprocity and optical isolation via coupling geometry.
(a): A design of directional coupler. Out of a small area on the turning, which is removed of cladding and close to the joint of two
fibers, the bending radiation from the field propagating along one direction hits the other fiber that is in a symmetric position with the first fiber. Meanwhile, the identical radiation due to the reversely propagating field mostly misses the other. (b): The consequent nonreciprocity of the individually acting fields. The gain rate $g_0$ with its saturation neglected is $0.9\gamma$, where $\gamma$ is the damping rate of the three cavities. Two pairs of CL and CCL modes in cavity 1 and cavity 3 couple at the rate $J_{13}\pm \epsilon$ with $J_{13}=10^{-2}\gamma$. Due to the directionality of bending radiation, the ratio $(J_{13}+ \epsilon)/(J_{13}-\epsilon)$ can be much higher than the displayed here, though the radiation's coupling into another fiber could have a low rate, i.e. $J_{13}\pm\epsilon\ll \gamma$ only given this type of coupling. The detuning $\Delta$ is zero for the coupling between fibers. (c1)-(c2): The improvement of the optical isolation with $J_{12}=J_{23}=2.2\gamma$ in (c1) to an approximate isolator operation with $J_{12}=J_{23}=10^{-2}\gamma$ in (c2). The other parameters are fixed as $\gamma_1=\gamma_2=\gamma$, $\gamma_3=9.72\gamma$, $g_0=9.7\gamma$, $J_{13}\pm \epsilon=(10^{-2}\pm 9\times 10^{-3})\gamma$, and $E_f=E_b=10\gamma$. In (c2) the isolation ratio defined in Eq. (3) is greatly enhanced together with the forward transmission proportional to $|a_{3,f}^-|$. }}
\vspace{-0.4cm}
\end{figure}

The above scenario provides a better understanding of the relation between Lorentz reciprocity and transmission nonreciprocity.
The scattering matrix is still symmetric after introducing the directional couplings in Fig. 4(a), 
but one of its additional symmetries is broken such that $S_{3-}^{1+}\neq S_{3+}^{1-}$. For two identical drives acting on $P_1$ and $P_4$ individually, 
such symmetry breaking results in the unequal forward output $\sqrt{2\kappa_e}a_{3,f}^-$ at $P_4$ and backward output $\sqrt{2\kappa_e}a_{1,b}^-$ at $P_1$, while the other pair of outputs, $\sqrt{2\kappa_e}a_{3,f}^+$ at $P_3$ and $\sqrt{2\kappa_e}a_{1,b}^+$ at $P_2$, constantly keep equal.
The relation $E_fa_{1,b}^+=E_b a_{3,f}^+$, as the manifestation of Lorentz reciprocity for the system due to its symmetric scattering matrix, 
is explicitly satisfied for any pair of forward input $E_f$ and backward input $E_b$; see Sec. III in \cite{supp} for more details. 
However, the nonreciprocal transmissions by applying the directional coupler in Fig. 4(a) to only two coupled cavities (cavity $2$ is detached) violate Lorentz reciprocity.

Such nonreciprocal transmissions by an MCCS are distinct from the asymmetric power flows in some other linear systems \cite{a1,l1}. 
For example, due to the drastically unequal transitions from one field mode to another mode with a different wave vector, asymmetric power flows between two ends exist in the system discussed in Refs. \cite{a1,a2}. But nonetheless, its intermodal transitions are identical, i.e. the transition from mode $1$ on the first end to mode $2$ on the second end has the same rate as that from mode $2$ on the second to mode $1$ on the first end. Corresponding to the circulating modes in MCCS, the forward transition from $a_1^+$ to $a_3^-$ is indeed symmetric with the backward one from $a_3^-$ to $a_1^+$, since the scattering matrix keeps symmetric. However, mode $a_3^-$ is from $P_3$ and mode $a_1^+$ leaves to $P_2$, not being the reversed transmission from $P_1$ to $P_4$. A field mode outputting to any port of such MCCS must be oppositely circulating with the mode that is excited by an input from the same port. At port $P_1$, for example, an input excites CL mode $a_1^+$, but any output comes from CCL mode $a_1^-$. The mutually reversed transmissions in Fig. 1(a), therefore, have to be the transitions from $a_1^+$ to $a_3^-$ and from $a_3^+$ to $a_1^-$, respectively. Such transitions from the CL (``$+$") to the CCL (``$-$") modes, unlike the intermodal ones mentioned above, will become nonidentical once the symmetry $S_{3-}^{1+}=S_{3+}^{1-}$ is broken, bringing about the transmission nonreciprocity.

In conclusion, we have demonstrated the nonreciprocal transmissions with an MCCS, illustrating such phenomena due to 
gain saturation and asymmetric field mode couplings. The structure exemplifies a new concept of realizing
optical nonreciprocity and optical isolation with the topology of its connected components and the geometry to couple the fields in different components, bearing fundamental dissimilarity with the other approaches that rely on direction-dependent processes in the used media or nonidentical couplings to the inputs from different ports. There also exist two distinct mechanisms for the phenomena: (1) to break Lorentz reciprocity simply with gain saturation, so that the inputs from different ports follow different dynamical equations; (2) to break the additional symmetries of the scattering matrix elements, with unequal couplings between pairs of circulating modes and/or unequal distances between pairs of cavities, while preserving the Lorentz reciprocity of the system at the same time. The nonreciprocal phenomena without breaking Lorentz reciprocity were previously thought to be impossible. By choosing the proper system parameters, the structure can work as an approximate optical isolator.

\begin{acknowledgements}
This work is supported in part by National Key R \& D program of China (Grant No. 2017YFA0303703) and NSFC (Grants
No. 11574093, 61435007 and 11574144). L.Y. is also sponsored by Fundamental Research Funds for the Central Universities
(Grant No. HEUCFJ170402) and NSFC (Grant No. 11747048).
\end{acknowledgements}


\end{document}